\documentclass[twocolumn,showpacs,aps,prl]{revtex4-1}
\usepackage{verbatim} %for comment
\usepackage{graphicx}% Include figure files
\usepackage{dcolumn}% Align table columns on decimal point
\usepackage{bm}% bold math
\usepackage{array}
\usepackage{booktabs}
\usepackage{amsmath}
\usepackage{color}
\usepackage{times}
\usepackage{epstopdf}
\usepackage{lineno}
\usepackage[T1]{fontenc}
\usepackage{ulem}
\begin{document}
\title{\boldmath
Measurement of cross sections for $e^{+}e^{-} \rightarrow \mu^+\mu^-$
at center-of-mass energies from 3.80 to 4.60 GeV
}
\author{
\begin{small}
\begin{center}
M.~Ablikim$^{1}$, M.~N.~Achasov$^{10,e}$, P.~Adlarson$^{63}$, S.~Ahmed$^{15}$, M.~Albrecht$^{4}$, A.~Amoroso$^{62A,62C}$, Q.~An$^{59,47}$,
~Anita$^{21}$, Y.~Bai$^{46}$, O.~Bakina$^{28}$, R.~Baldini Ferroli$^{23A}$,
I.~Balossino$^{24A}$, Y.~Ban$^{37,m}$, K.~Begzsuren$^{26}$,
J.~V.~Bennett$^{5}$,
M.~Bertani$^{23A}$,
D.~Bettoni$^{24A}$, F.~Bianchi$^{62A,62C}$, J~Biernat$^{63}$,
J.~Bloms$^{56}$, A.~Bortone$^{62A,62C}$, I.~Boyko$^{28}$,
R.~A.~Briere$^{5}$, H.~Cai$^{64}$, X.~Cai$^{1,47}$, A.~Calcaterra$^{23A}$,
G.~F.~Cao$^{1,51}$, N.~Cao$^{1,51}$, S.~A.~Cetin$^{50B}$,
J.~F.~Chang$^{1,47}$, W.~L.~Chang$^{1,51}$, G.~Chelkov$^{28,c,d}$,
D.~Y.~Chen$^{6}$, G.~Chen$^{1}$, H.~S.~Chen$^{1,51}$, M.~L.~Chen$^{1,47}$,
S.~J.~Chen$^{35}$, X.~R.~Chen$^{25}$, Y.~B.~Chen$^{1,47}$, W.~Cheng$^{62C}$,
G.~Cibinetto$^{24A}$, F.~Cossio$^{62C}$, X.~F.~Cui$^{36}$,
H.~L.~Dai$^{1,47}$, J.~P.~Dai$^{41,i}$, X.~C.~Dai$^{1,51}$,
A.~Dbeyssi$^{15}$, R.~ B.~de Boer$^{4}$, D.~Dedovich$^{28}$,
Z.~Y.~Deng$^{1}$, A.~Denig$^{27}$, I.~Denysenko$^{28}$,
M.~Destefanis$^{62A,62C}$, F.~De~Mori$^{62A,62C}$, Y.~Ding$^{33}$,
C.~Dong$^{36}$, J.~Dong$^{1,47}$, L.~Y.~Dong$^{1,51}$,
M.~Y.~Dong$^{1,47,51}$, S.~X.~Du$^{67}$, J.~Fang$^{1,47}$,
S.~S.~Fang$^{1,51}$, Y.~Fang$^{1}$, R.~Farinelli$^{4A,24B}$,
L.~Fava$^{62B,62C}$, F.~Feldbauer$^{4}$, G.~Felici$^{23A}$,
C.~Q.~Feng$^{59,47}$, M.~Fritsch$^{4}$, C.~D.~Fu$^{1}$, Y.~Fu$^{1}$,
X.~L.~Gao$^{59,47}$, Y.~Gao$^{60}$, Y.~Gao$^{37,m}$, Y.~G.~Gao$^{6}$,
I.~Garzia$^{24A,24B}$, E.~M.~Gersabeck$^{2020}$,
K.~Goetzen$^{11}$, L.~Gong$^{36}$, W.~X.~Gong$^{1,47}$, W.~Gradl$^{27}$,
M.~Greco$^{62A,62C}$, L.~M.~Gu$^{35}$, M.~H.~Gu$^{1,47}$, S.~Gu$^{2}$,
Y.~T.~Gu$^{13}$, C.~Y~Guan$^{1,51}$
% ...
\\
\vspace{0.2cm}
(BESIII Collaboration)\\
\vspace{0.2cm} {\it
$^{1}$ Institute of High Energy Physics, Beijing 100049, People's Republic
of China\\
$^{2}$ Beihang University, Beijing 100191, People's Republic of China\\
% ...
$^{4}$ Bochum Ruhr-University, D-44780 Bochum, Germany\\
$^{5}$ Carnegie Mellon University, Pittsburgh, Pennsylvania 15213, USA\\
$^{6}$ Central China Normal University, Wuhan 430079, People's Republic of
China\\
% ...
$^{10}$ G.I. Budker Institute of Nuclear Physics SB RAS (BINP), Novosibirsk
630090, Russia\\
$^{11}$ GSI Helmholtzcentre for Heavy Ion Research GmbH, D-64291 Darmstadt,
Germany\\
% ...
$^{13}$ Guangxi University, Nanning 530004, People's Republic of China\\
% ...
$^{15}$ Helmholtz Institute Mainz, Johann-Joachim-Becher-Weg 45, D-55099
Mainz, Germany\\
% ...
$^{21}$ Indian Institute of Technology Madras, Chennai 600036, India\\
% ...
$^{23}$ (A)INFN Laboratori Nazionali di Frascati, I-00044, Frascati, Italy;
(B)INFN and University of Perugia, I-06100, Perugia, Italy\\
$^{24}$ (A)INFN Sezione di Ferrara, I-44122, Ferrara, Italy; (B)University
of Ferrara, I-44122, Ferrara, Italy\\
$^{25}$ Institute of Modern Physics, Lanzhou 730000, People's Republic of
China\\
$^{26}$ Institute of Physics and Technology, Peace Ave. 54B, Ulaanbaatar
13330, Mongolia\\
$^{27}$ Johannes Gutenberg University of Mainz, Johann-Joachim-Becher-Weg
45, D-55099 Mainz, Germany\\
$^{28}$ Joint Institute for Nuclear Research, 141980 Dubna, Moscow region,
Russia\\
% ...
$^{33}$ Liaoning University, Shenyang 110036, People's Republic of China\\
% ...
$^{35}$ Nanjing University, Nanjing 210093, People's Republic of China\\
$^{36}$ Nankai University, Tianjin 300071, People's Republic of China\\
$^{37}$ Peking University, Beijing 100871, People's Republic of China\\
% ...
$^{41}$ Shanghai Jiao Tong University, Shanghai 200240, People's Republic of
China\\
% ...
$^{46}$ Southeast University, Nanjing 211100, People's Republic of China\\
$^{47}$ State Key Laboratory of Particle Detection and Electronics, Beijing
100049, Hefei 230026, People's Republic of China\\
% ...
$^{50}$ (A)Ankara University, 06100 Tandogan, Ankara, Turkey; (B)Istanbul
Bilgi University, 34060 Eyup, Istanbul, Turkey; (C)Uludag University, 16059
Bursa, Turkey; (D)Near East University, Nicosia, North Cyprus, Mersin 10,
Turkey\\
$^{51}$ University of Chinese Academy of Sciences, Beijing 100049, People's
Republic of China\\
% ...
$^{56}$ University of Muenster, Wilhelm-Klemm-Str. 9, 48149 Muenster,
Germany\\
% ...
$^{59}$ University of Science and Technology of China, Hefei 230026,
People's Republic of China\\
$^{60}$ University of South China, Hengyang 421001, People's Republic of
China\\
% ...
$^{62}$ (A)University of Turin, I-10125, Turin, Italy; (B)University of
Eastern Piedmont, I-15121, Alessandria, Italy; (C)INFN, I-10125, Turin,
Italy\\
$^{63}$ Uppsala University, Box 516, SE-75120 Uppsala, Sweden\\
$^{64}$ Wuhan University, Wuhan 430072, People's Republic of China\\
% ...
$^{67}$ Zhengzhou University, Zhengzhou 450001, People's Republic of China\\
\vspace{0.2cm}
% ...
$^{c}$ Also at the Moscow Institute of Physics and Technology, Moscow
141700, Russia\\
$^{d}$ Also at the Functional Electronics Laboratory, Tomsk State
University, Tomsk, 634050, Russia\\
$^{e}$ Also at the Novosibirsk State University, Novosibirsk, 630090,
Russia\\
% ...
$^{i}$ Also at Key Laboratory for Particle Physics, Astrophysics and
Cosmology, Ministry of Education; Shanghai Key Laboratory for Particle
Physics and Cosmology; Institute of Nuclear and Particle Physics, Shanghai
200240, People's Republic of China\\
% ...
$^{m}$ Also at State Key Laboratory of Nuclear Physics and Technology,
Peking University, Beijing 100871, People's Republic of China\\
% ...
      }
    \end{center}
    \vspace{0.4cm}
  \end{small}
}
\noaffiliation{}

\vspace{0.0cm}
%\date{\today}

\begin{abstract}
The observed cross sections for $e^+e^-\rightarrow \mu^+\mu^-$ at
energies from 3.8 to 4.6 GeV are measured using data samples taken with the
BESIII detector operated at the BEPCII collider.
We measure the muonic widths and determine the branching fractions
of the charmonium states $\psi(4040)$, $\psi(4160)$, and $\psi(4415)$ decaying to $\mu^+\mu^-$,
as well as making a first determination of
the phase of the amplitudes.
In addition, we observe evidence for a structure
in the dimuon cross section near 4.220 GeV/$c^2$,
which we denote as $S(4220)$.
Analyzing a coherent sum of amplitudes yields
eight solutions, one of which gives
a mass of ${M}_{S(4220)}=4216.7 \pm 8.9 \pm 4.1$~MeV/$c^2$,
a total width of ${\rm \Gamma^{\rm tot}_{S(4220)}}=47.2 \pm 22.8 \pm 10.5$~MeV,
and a muonic width of ${\rm \Gamma}^{\mu\mu}_{S(4220)}=1.53\pm1.26\pm0.54$~keV,
where the first uncertainties are statistical and the second systematic.
The eight solutions give the central values of the mass,
total width, muonic width to be, respectively, in the range
from 4212.8 to 4219.4 MeV/$c^2$, from 36.4 to 49.6 MeV, and from 1.09 to 1.53 keV.
The statistical significance of the $S(4220)$ signal is $3.9\sigma$.
Correcting the total dimuon cross section
for radiative effects yields a statistical
significance for this structure of more than
$7\sigma$.
\end{abstract}
%
%%\pacs{   }% PACS

\maketitle

\vspace{0.0cm}

For a long time the meson resonances produced in ${\rm e^+e^-}$ collisions
above the open-charm (OC) and below the open-bottom thresholds
had been thought to decay entirely to OC final states
through the strong interaction.
Consequently, the possibility of non-open-charm (NOC) decays
attracted little experimental interest until the early
years of the millennium.
For convenience, in this Letter we denote these resonances
$X_{\rm {above~D{\bar D}}}$,
which encompasses both heavy $c {\bar c}$ states, \emph{i.e.}
$\psi(3770)$, $\psi(4040)$, $\psi(4160)$,
and $\psi(4415)$, and non-$c {\bar c}$ states, such as
four-quark composites, hybrid charmonium states,
and open-charm molecule states~\cite{
F_E_Close_et_al_PLB578_119_y2004,
M_B_Voloshin_JETP_Lett_23_333_y1976,
HQ_ProgressPuzzleOpportunities_EurPhysJ_C72_1534_Year2011} that are expected by QCD.
Finding these non-$c\bar c$ states
would be a crucial validation of the QCD predictions.

Since non-$c {\bar c}$ states may easily decay to non-open-charm final states,
such decays of $X_{\rm {above~D{\bar D}}}$ mesons
were searched for by the BES collaboration
using the data collected with
the BES-I detector at energies of 4.04 and 4.14 GeV,
and the BES-II detector at energies around 3.773 GeV.
The first evidence for such decays
was reported in the $J/\psi\pi^+\pi^-$ final state by BES in 2003 \cite{hep_ex_0307028v1}.
This final state could come from the decay of a $c{\bar c}$
or a non-$c {\bar c}$ state, or even both of these states.
On the assumption that there is no other resonance
at energies near 3.773 GeV, the signal
was interpreted to be $\psi(3770) \rightarrow J/\psi\pi^+\pi^-$~\cite{PLB605_Y2005_63}.
This first NOC decay was confirmed by the CLEO collaboration~\cite{CLEO_JPisPiPi}
two year after the BES analysis.
This discovery overturned the conventional understanding that
${\rm X}_{\rm above~D{\bar D}}$ decay into open-charm final states
through the strong interaction with branching fractions of almost 100\%.
It stimulated strong interest in searching for other non-open-charm decays
of $X_{\rm {above~D{\bar D}}}$ mesons~\cite{BES2_Psi3770_Program}, in particular into
$J/\psi\pi^+\pi^-$ and similar final states,  and
led to the discovery of several new
resonances~\cite{X3872_PhysRevLett91_262001_Y2003, X4260_PRL95_142001_Y2005,X4360andX4660_PRL99_142002_Y2007}.

In the last 17 years, several new
states~\cite{X3872_PhysRevLett91_262001_Y2003, X4260_PRL95_142001_Y2005,
X4360andX4660_PRL99_142002_Y2007}, and
new di-structures, such as the  $Rs(3770)$~\cite{bes2_prl_2structures} and
$R(4220)$ and $R(4320)$~\cite{PRL118_092001_2017},
as well as structures lying above 4.2\,GeV~\cite{PRL114_092993_2015,Y4220_BESIII_PRL118_092002_2017}
have been observed in $e^+e^-$ annihilation at
energies above the open-charm threshold.  The $X(3872)$~\cite{X3872_PhysRevLett91_262001_Y2003},
$Y(4260)$~\cite{X4260_PRL95_142001_Y2005},
and $R(4220)$ and $R(4320)$~\cite{PRL118_092001_2017}
resonances were observed in the $J/\psi\pi^+\pi^-$ final state,
while the $Y(4360)$~\cite{X4360andX4660_PRL99_142002_Y2007}
and $Y(4660)$\cite{X4360andX4660_PRL99_142002_Y2007}
were observed in the $\psi(3686)\pi^+\pi^-$ final state.
In addition, the $Y(4220)$~\cite{PRL114_092993_2015} was observed
in the final state $\omega\chi_{\rm c0}$, and the 
$Y(4220)$ and $Y(4390)$~\cite{PRL118_092001_2017} were observed 
in the final state ${h_c}\pi^+\pi^-$. All of these resonances were
observed in final states of inclusive hadrons,
where no attempt was made to identify the hadron species,
and in final states of ${M_{c\bar c}X_{\rm LH}}$,
where ${M}_{c\bar c}$ is a hidden-charm meson
and ${X_{\rm LH}}$ is a light hadron.
In Ref.~\cite{RongG_CPC_34_778_Y2010} it was suggested
to search for new vector
states at BESIII by means of analyzing the line-shape of cross sections for
$e^+e^-\rightarrow f_{\rm NOC}$, where $f_{\rm NOC}$ can be
$J/\psi X$, $\psi(3686) X$ ($X$ =light hadrons or photons), light hadrons only, or inclusive hadrons.

In addition to studying the final states
$J/\psi X$ or $\psi(3686) X$ produced in $e^+e^-$ annihilations,
searches for new vector states may be performed by analyzing the cross
section for $e^+e^- \rightarrow \mu^+\mu^-$, in which
the contribution from the decays of heavy $c{\bar c}$ resonances
are strongly suppressed,
and consequently the production and decay of the non-$c{\bar c}$ states
can be significantly enhanced.

In this Letter, we report measurements of the cross section for
$e^+e^-\rightarrow \mu^+\mu^-$ at center-of-mass (c.m.) energies from 3.8 to 4.6~GeV, 
and studies of the known $c\bar{c}$ resonances and searches for new structures in this regime by
performing an analysis of a coherent sum of amplitudes contributing
of this cross section.
The data samples used in measuring the cross section
were collected at 133 c.m.\ energies with the BESIII detector
operated at the BEPCII collider from 2011 to 2017.
The total integrated luminosity of the data sets used in the analysis
is 13.2~fb$^{-1}$, determined from  large-angle
Bhabha scattering events~\cite{LumOfTheData}.
The c.m.\ energy of each data set is measured using dimuon events, with
an uncertainty of $\pm 0.8$ MeV~\cite{Ecm_Uncertainty_Published}.

The BESIII detector is described in detail in Ref.~\cite{bes3}.
The detector response is studied using samples of Monte Carlo (MC)
events which are simulated with the {\sc Geant4}-based~\cite{geant4}
detector simulation software package {\sc boost}.
Simulated samples for all vector $q\bar q$ states (i.e. $u\bar u$, $d\bar d$,
$s\bar s$, and $c \bar c$ resonances) and their decays to
$\mu^+\mu^-$ are generated with the MC event generator
{\sc BabaYaga}~\cite{babayaga}.
Possible background sources are estimated with Monte Carlo simulated
events generated with the event generator
{\sc kkmc}~\cite{kkmc}.
The detection efficiency is determined with Monte Carlo simulated
$e^+e^-\rightarrow \mu^+\mu^-$ events generated with {\sc BabaYaga},
which includes initial and final state radiation, as well as vacuum polarization effects.

The observed cross section for $e^+e^- \rightarrow \mu^+ \mu^-$ at a
certain c.m.\ energy $\sqrt{s}$ is determined by
\begin{linenomath*}
\begin{equation}
   \sigma^{\rm obs}(e^+e^-\rightarrow \mu^+\mu^-)
    =\frac{N^{\rm obs} }
   {\mathcal L \epsilon
   },
   \label{equation:Eq_ObsCS}
\end{equation}
\end{linenomath*}
\noindent
where $N^{\rm obs}$ is the background-subtracted number of observed events
for $e^+e^-\rightarrow \mu^+\mu^-$,
$\mathcal L$ is the integrated luminosity,
and $\epsilon$ is the detection efficiency.

Each candidate for $e^+e^-\rightarrow \mu^+\mu^-$ is required to have two tracks of opposite charge.
Each of the two charged tracks must satisfy $|\cos\theta|<0.8$,
where $\theta$ is the polar angle of the  tracks.
In addition, the charged tracks are required to satisfy
$V_r<1.0$~cm and $|V_z|<10.0$~cm, where $V_r$ is
the distance of closest approach to the interaction point
in the $r$-$\phi$
plane, and $|V_z|$ is the distance between the point of the closest approach
and the interaction point
along the beam axis.
Furthermore, the total momentum
$|\vec p_+| + |\vec p_-|$
of the two charged tracks
is required to be greater than 90\% of the
nominal collision energy $\sqrt{s}$.
To reject Bhabha scattering events, we require the ratio
of the energy $E_{\pm}$ deposited in the electromagnetic
calorimeter to the momentum $p_{\pm}$ of the charged track
to satisfy  $0.05<E_{\pm}/p_{\pm}<0.40$.
This criterion also rejects $\pi^+\pi^-$ pairs.
The rejection fraction for $\pi^+\pi^-$ events is energy dependent,
ranging from $41.5\%$ at 3.8~GeV to $37.5\%$ at 4.6~GeV.
The remaining $\pi^+\pi^-$ background is subtracted using the
extrapolation of the $e^+e^-\rightarrow \pi^+\pi^-$ cross section
measured by the CLEO collaboration~\cite{CLEO_PhysRevLett95_261803_2005}
and the rate of misidentifying $\pi^+\pi^-$ as $\mu^+\mu^-$
obtained from the MC simulation.
In order to reduce the $K^+K^-$ and $p\bar p$ background,
the event is subjected to a four-constraint kinematic fit with the hypothesis
$e^+e^-\rightarrow \mu^+\mu^-$, constraining the total four-momentum of the
$\mu^+\mu^-$ to that of the colliding beams, and
the fit $\chi^2_{\rm 4C}$ is required to be less than 60.

\begin{figure}[]
\centering
\includegraphics[width=0.495\textwidth]{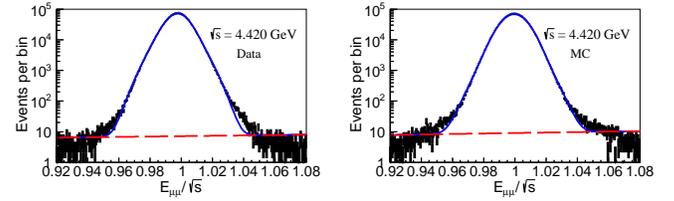}
  \caption{
Distributions of the ratios of the total energies
$E_{\mu^+\mu^-}$ of the ${\mu^+\mu^-}$ system
to $\sqrt{s}$
for the events selected from the data (left) collected at $\sqrt{s}=4.42$ GeV
and MC events (right) simulated at the same energy.
The black dots with error bars show the
data and MC events, the blue solid line shows the fit to these, and the red dashed line shows the backgrounds.
  }
 \label{fig:Emumu_over_Ecm_ecm4420MeV}
\end{figure}

The number of $e^+e^-\rightarrow \mu^+\mu^-$ candidates is
determined by analyzing the ratio $E_{\mu^+\mu^-}/\sqrt{s}$, where
$E_{\mu^+\mu^-}$ is the total energy of $\mu^+$ and $\mu^-$ determined
from the measured track momenta.
As an example,
Fig.~\ref{fig:Emumu_over_Ecm_ecm4420MeV} (left)
shows the distribution of $E_{\mu^+\mu^-}/\sqrt{s}$
for the events selected from the data
collected at $\sqrt{s}=4.420$\,GeV.
A fit to the distribution with a double-Gaussian function for the signal
shape and a first order polynomial to describe the background
yields $N^{\rm fit}$=$(2500.2\pm1.6)\times 10^{3}$
$e^+e^-\rightarrow \mu^+\mu^-$ candidates, where the uncertainty
is statistical. The systematic uncertainty due to the
non-peaking background
(mainly cosmic rays and beam-gas events)
is estimated to be less than 0.01\%, and therefore negligible.
The imperfection of the signal peak description
is taken into account as a systematic uncertainty (see below).
The signal yield $N^{\rm fit}$ is still
contaminated by peaking background from several sources, e.g.
$e^+e^-\rightarrow (\gamma)e^+e^-$,
$e^+e^-\rightarrow \pi^+\pi^-$, and
$e^+e^-\rightarrow K^+K^-$.
Using the high-statistics samples of MC simulated events and the extrapolated
cross sections for these processes,
the number of the background events is estimated to be $N^{\rm b}=4764\pm 18$,
where the uncertainty is mostly due to the cross-section extrapolation.
Subtracting $N^{\rm b}$ from
$N^{\rm fit}$ yields $N^{\rm obs}$=$(2495.4\pm 1.6)\times 10^{3}$
signal events.

The integrated luminosity of the data sample taken at 4.420~GeV was previously
measured to be $\mathcal L=1043.9\pm 0.1 \pm 6.9$
pb$^{-1}$~\cite{LumOfTheData}, where the first uncertainty is
statistical and the second one is systematic.
At 4.420 GeV, the detection efficiency
of $e^+e^-\rightarrow \mu^+\mu^-$ is
$\epsilon=(41.09\pm 0.01)\%$, as determined from the MC.
Using these values in
Eq.~(\ref{equation:Eq_ObsCS})
yields the observed cross section of
$\sigma^{\rm obs}(e^+e^- \rightarrow \mu^+\mu^-)=5.818\pm 0.010 \pm 0.169$~nb.
The first error includes the uncertainties
of statistical origin (signal sample size,
MC event statistics and the statistical uncertainty
of the luminosity measurement).
The second error represents the remaining systematic
uncertainties (see below).
Similarly, we determine the observed cross sections
for $e^+e^- \rightarrow \mu^+\mu^-$ at the other 132 energies
from 3.81 to 4.6 GeV.

The systematic uncertainty for the observed cross section originates
from several sources. They are
$1\%$ due to the luminosity measurement,
$1\%$ per track associated with the knowledge of the tracking efficiency,
$0.64\%$ due to requiring $|\cos\theta| < 0.8$,
$0.59\%$ due to requiring
$|\vec p_+| + |\vec p_-|>0.9\sqrt{s}$,
$0.12\%$ due to the selection on $E_{\pm}/|\vec p_{\pm}|$,
$0.41\%$ due to the four-constraint kinematic fit,
$1.23\%$ due to the fit to the $E_{\mu^+\mu^-}/\sqrt{s}$ distribution,
$0.03\%$ due to the background subtraction,
and
$1\%$ due to the theoretical uncertainty associated with the Monte Carlo generator.
An additional uncertainty arises from the imperfect
description of the signal shape by the fit
(see Figure \ref{fig:Emumu_over_Ecm_ecm4420MeV}).
This effect is only partially compensated by the MC,
and the residual uncertainty is 0.03\%.
Adding these uncertainties in quadrature yields
a total systematic uncertainty of $2.91\%$.

To search for new vector states in
$e^+e^- \rightarrow \mu^+ \mu^-$, a $\chi^2$ fit is performed to the
measured cross section.
In the fit, the expected cross section is given by~\cite{PLB641_Y2006_145,PRL97_262001_Y2006}
\begin{eqnarray}
  \sigma^{\rm exp}_{{\mu^+}{\mu^-}}(s)= \int^{ 1- \frac{4m^2_{\mu}}{s} }_0  dx \cdot
        {\sigma^{\rm D}_{\mu^+\mu^-}(s(1-x))} F(x,s),
 \label{equation:Expected_CrossSection}
\end{eqnarray}
\noindent
where $m_{\mu}$ is the mass of muon and
$F(x,s)$ is a sampling function~\cite{Structure_Function}
for the radiative photon energy fraction $x$.
$\sigma^{\rm D}(s(1-x))$ is the dressed cross section
including vacuum-polarization effects,
\begin{eqnarray}
\sigma^{\rm D}_{\mu^+\mu^-}(s(1-x)) =
\left|A_{\rm cnt} +
\sum\limits_{k=1}^{9} e^{i\phi_{R_k}}A_{R_k} + e^{i\phi_{S}}A_{S}
   \right|^2
 \label{equation:Drss_CrossSection_A}
\end{eqnarray}
where $A_{\rm cnt}$, $A_{R_k}$ and $A_{\rm S}$ are, respectively,
the amplitude for continuum $e^+e^-\rightarrow \mu^+\mu^-$ production,
the Breit-Wigner (BW) amplitude describing nine vector resonances
($\rho(770)$, $\omega(782)$, $\phi(1020)$, $J/\psi$, $\psi(3686)$, $\psi(3770)$,
$\psi(4040)$, $\psi(4160)$, and $\psi(4415)$), and a new vector structure $S$
decaying into $\mu^+\mu^-$,
while $\phi_{R_k}$ and $\phi_{S}$ are the corresponding phases of the
amplitudes.
The continuum amplitude
can be parameterized as
$A_{\rm cnt} =\sqrt{f_{\rm cnt}/{s'}}$,
where $f_{\rm cnt}$ is a free parameter, and $s'=s(1-x)$.
The decay amplitude of resonance $\mathcal R$, being either one of the
known vector states or the new structure $S$, is written as
${\mathcal A} =
 \sqrt{12\pi
 \Gamma^{ee}_{\mathcal R}\Gamma^{\mu\mu}_{\mathcal R} }
      /[{(s'-M_{\mathcal R}^2) + i \Gamma^{\rm tot}_{\mathcal R}M_{\mathcal R}}]$,
where $M_{\mathcal R}$, $\Gamma^{ee}_{\mathcal R}$, $\Gamma^{\mu\mu}_{\mathcal R}$
and $\Gamma^{\rm tot}_{\mathcal R}$ are the mass,
electron width, muonic width, and
total width, respectively.

In the fit the observed cross-section values
are assumed to be influenced only by the
uncertainties of statistical origin.
The uncertainties on the parameters returned by the fit
are referred to as statistical uncertainties
in the subsequent discussion.
The remaining cross-section
uncertainties (assumed to be fully correlated
between different energies) are taken into account
using the ``offset method'' \cite{offset}:
the cross-section values are changed
for all energies simultaneously by the
size of the uncertainty and the resulting change
in the fit parameter is taken
as a systematic uncertainty.

Since the analysis does not include the observed cross section at energies
below 3.8 GeV, the parameters of the
first six lower mass vector resonances are all fixed to the values given
by the particle data group (PDG)~\cite{PDG2018},
and the phases are fixed to zero.
For the three heavy $c\bar c$ states, \emph{i.e.} $\psi(4040)$, $\psi(4160)$, and
$\psi(4415)$, the masses and the total widths are also fixed to the values
given by the PDG.
The partial widths  $\Gamma^{\mu\mu}$ and the phases are left free, and lepton universality is assumed
(i.e. $\Gamma^{ee}_{\mathcal R}=\Gamma^{\mu\mu}_{\mathcal R}$). It is noted that the earlier studies contributing
to the values for $\Gamma^{ee}_{\mathcal R}$ reported in Ref.~\cite{PDG2018} did not consider the contributions
from non-$c\bar{c}$ states in the calculations of the
Initial State Radiative (ISR)
correction factors; furthermore they assumed a selection efficiency
for  $e^+e^-\rightarrow {\rm hadrons}$ that is a smooth curve, increasing as the c.m.\ energy increases,
rather than the BW-like function observed in \emph{e.g.} Fig. 1(b) of Ref.~\cite{BES2_PRL97_121801_Y2006}.
Neglecting these effects may lead to bias,
as may the difficulties of accounting for interference effects between the
continuum $e^+e^-\rightarrow {\rm hadrons}$ amplitude and the resonance decay
amplitudes.
Following these considerations we leave these partial widths as free parameters in our fit.

The fit returns eight acceptable solutions with distinct results  for the four free phases.
Table~\ref{tab:FitResults_CntPsi3686S37xxP3770P4040P4160S4230P4415_Solution_ABCD_EFGH}
shows the results from the fit.  All solutions include a result for a new structure with mass close to 4220~MeV,
and so we denote this possible state as  $S(4220)$.
For Solution I, the fit returns
$f_{\rm cnt}=88.51\pm 0.11$ nb/GeV$^2$
and
$\chi^2=135.47$ for 121 degrees of freedom.
Taking
$\Gamma^{\mu\mu}_{S(4220)}=\Gamma^{\rm tot}_{S(4220)}{\mathcal B}(S(4220)\rightarrow\mu^+\mu^-)$,
where ${\mathcal B}(S(4220)\rightarrow\mu^+\mu^-)$ is the branching fraction for the decay of
$S(4220)\rightarrow\mu^+\mu^-$,
the fit yields
$\Gamma^{ee}_{S(4220)}{\mathcal B}(S(4220)\rightarrow \mu^+\mu^-)=0.05\pm0.06\pm 0.03$ eV,
where the first uncertainty  is statistical and the second is systematic.
\begin{table*}
\centering
\caption{Results from the fit to the $e^+e^-\rightarrow \mu^+\mu^-$  cross section showing the values of the muonic width $\Gamma^{\mu\mu}_{\mathcal R_i}$ [in keV], branching fraction $\mathcal B(\mathcal R_i \rightarrow \mu^+ \mu^-)$
and phase $\phi_{\mathcal R_i}$ [in degrees], where $\mathcal R_1$, $\mathcal R_2$, $\mathcal R_3$ and $\mathcal R_4$
represent $\psi(4040)$, $\psi(4160)$, $\psi(4415)$ and $S(4220)$, respectively.
Also given is the mass  $M_{\mathcal R_4}$ [in MeV],
and total width $\Gamma^{\rm tot}_{\mathcal R_4}$ of the $S(4220)$.
The first uncertainties are statistical, and the second
are systematic.}
\begin{tabular}{lcccr}
\hline\hline
Parameters       & Solution I      &  Solution II      &  Solution III  &    Solution IV  \\
\hline
$\Gamma^{\mu\mu}_{\mathcal R_1}$
    &  $0.73 \pm 0.48 \pm 0.12$  & $0.62 \pm 0.46 \pm 0.10$ & $0.61 \pm 0.46 \pm 0.10$ & $0.71 \pm 0.42 \pm 0.12$  \\
$\mathcal B(\mathcal R_1 \rightarrow \mu^+ \mu^-)$
    & $0.91 \pm 0.60 \pm 0.20$   & $0.77 \pm 0.58 \pm 0.17$ & $0.76 \pm 0.58 \pm 0.16$ & $0.89 \pm 0.53 \pm 0.19$    \\
$\phi_{\mathcal R_1}$
    & $-77  \pm 33   \pm 11$     & $-75  \pm 38   \pm 11$   & $-76  \pm 1    \pm 11$   &  $-77  \pm 39   \pm 11$     \\
$\Gamma^{\mu\mu}_{\mathcal R_2}$
    & $2.45 \pm 1.24 \pm 0.94$   & $2.36 \pm 1.26 \pm 0.91$ & $2.31 \pm 1.08 \pm 0.89$   & $2.41 \pm 1.08 \pm 0.93$     \\
$\mathcal B(\mathcal R_2 \rightarrow \mu^+ \mu^-)$
    & $3.49 \pm 1.78 \pm 1.22$   & $3.37 \pm 1.80 \pm 1.18$   & $3.30 \pm 1.55 \pm 1.16$   & $3.45 \pm 1.54 \pm 1.21$  \\
$\phi_{\mathcal R_2}$
    & $153  \pm 33   \pm 11$   & $138  \pm 29   \pm 10$     & $137  \pm 29   \pm 10$    &  $150  \pm 11   \pm 11$         \\
$\Gamma^{\mu\mu}_{\mathcal R_3}$
    & $1.25 \pm 0.28 \pm 0.35$  & $1.26 \pm 0.27 \pm 0.35$   & $1.27 \pm 0.27 \pm 0.36$  & $1.24 \pm 0.27 \pm 0.35$     \\
$\mathcal B(\mathcal R_3 \rightarrow \mu^+ \mu^-)$
    & $2.01 \pm 0.44 \pm 0.87$  & $2.03 \pm 0.44 \pm 0.88$ & $2.05 \pm 0.44 \pm 0.89$   & $2.01 \pm 0.44  \pm 0.87$      \\
$\phi_{\mathcal R_3}$
    & $-26  \pm 13   \pm 10$    & $-28  \pm 13   \pm 11$   & $-28  \pm 12   \pm 11$    & $-27  \pm 12    \pm 11$          \\
%%%%%  &                             &                           &                           &                                   \\
${\rm M}_{\mathcal R_4}$
    & $4216.7 \pm 8.9  \pm 4.1$   & $4213.6 \pm 7.5  \pm 4.1$   & $4213.5 \pm 7.4  \pm 4.1$   &  $4216.2 \pm 5.7  \pm 4.1$            \\
${\rm \Gamma^{\rm tot}_{\mathcal R_4}}$
    & $47.2 \pm 22.8 \pm 10.5$   & $39.9 \pm 19.5 \pm 8.9$    & $38.8 \pm 17.5 \pm 8.6$   & $45.5 \pm 13.3 \pm 10.1$  \\
$\Gamma^{\mu\mu}_{\mathcal R_4}$
    &$1.53 \pm 1.26 \pm 0.54$   &  $1.28 \pm 1.09 \pm 0.46$ &  $1.22 \pm 0.93 \pm 0.43$ &  $1.46 \pm 0.89 \pm 0.52$   \\
$\phi_{\mathcal R_4}$
    & $20   \pm 44   \pm 13$     & $0     \pm 40   \pm 0$     &  $-2   \pm 39   \pm 1$    &  $17   \pm 19   \pm 11$     \\
\hline
Parameters
      & Solution V             &  Solution VI               &  Solution VII             & Solution   VIII  \\
\hline
$\Gamma^{\mu\mu}_{\mathcal R_1}$
   & $0.74 \pm 0.50 \pm 0.12$   & $0.58 \pm 0.46 \pm 0.10$  &  $0.66 \pm 0.46 \pm 0.11$   & $0.80 \pm 0.48 \pm 0.13$      \\
$\mathcal B(\mathcal R_1 \rightarrow \mu^+ \mu^-)$
   & $0.93 \pm 0.63 \pm 0.20$      & $0.72 \pm 0.57 \pm 0.16$      &  $0.83 \pm 0.58 \pm 0.18$ & $1.00 \pm 0.60 \pm 0.22$    \\
$\phi_{\mathcal R_1}$
   &  $-78  \pm 31   \pm 11$       & $-73  \pm 41   \pm 10$        &  $-69  \pm 37   \pm 10$   & $284  \pm 30   \pm 40$     \\
$\Gamma^{\mu\mu}_{\mathcal R_2}$
   &  $2.28 \pm 1.05 \pm 0.88$     & $2.22 \pm 1.12 \pm 0.85$ &  $2.08 \pm 0.99 \pm 0.80$   & $2.31 \pm 1.14 \pm 0.89$  \\
$\mathcal B(\mathcal R_2 \rightarrow \mu^+ \mu^-)$
    & $3.26 \pm 1.46 \pm 1.14$     & $3.17 \pm 1.61 \pm 1.11$ & $2.97 \pm 1.41 \pm 1.04$       & $3.29 \pm 1.63 \pm 1.15$ \\
$\phi_{\mathcal R_2}$
    & $157  \pm 37   \pm 12$       & $132  \pm 28   \pm 10$      &  $143  \pm 29   \pm 11$        & $154  \pm 31   \pm 12$          \\
$\Gamma^{\mu\mu}_{\mathcal R_3}$
    & $1.24 \pm 0.28 \pm 0.35$  & $1.27 \pm 0.27 \pm 0.36$   &  $1.24 \pm 0.27 \pm 0.35$   & $1.25 \pm 0.27 \pm 0.35$  \\
$\mathcal B(\mathcal R_3 \rightarrow \mu^+ \mu^-)$
    & $2.00 \pm 0.45 \pm 0.87$     & $2.05 \pm 0.43 \pm 0.89$ & $2.01 \pm 0.43 \pm 0.87$       & $2.02 \pm 0.44 \pm 0.87$      \\
$\phi_{\mathcal R_3}$
    & $-25  \pm 13    \pm 10$     & $-29  \pm 12   \pm 11$       & $-28  \pm 13   \pm 11$      & $332  \pm 13   \pm 132$     \\
${\rm M}_{\mathcal R_4}$
    & $4219.4 \pm 11.2 \pm 4.1$  & $4212.8 \pm 7.2  \pm 4.0$   &  $4216.1 \pm 7.5  \pm 4.1$   & $4217.3 \pm 9.1  \pm 4.1$   \\
${\rm \Gamma^{\rm tot}_{\mathcal R_4}}$
    & $49.6 \pm 22.6 \pm 11.0$   & $36.4 \pm 16.8 \pm 8.1$    &  $37.8 \pm 18.5 \pm 8.4$    & $45.5 \pm 21.2 \pm 10.1$   \\
$\Gamma^{\mu\mu}_{\mathcal R_4}$
   & $1.50 \pm 1.03 \pm 0.53$  &  $1.12 \pm 0.89 \pm 0.40$ &  $1.09 \pm 0.84 \pm 0.39$  &  $1.40 \pm 1.08 \pm 0.50$ \\
$\phi_{\mathcal R_4}$
    & $31   \pm 51   \pm 20$   & $-8   \pm 39   \pm 5$          & $10   \pm 40   \pm 7$ &  $22   \pm 44   \pm 15$   \\
\hline\hline
\end{tabular}
\label{tab:FitResults_CntPsi3686S37xxP3770P4040P4160S4230P4415_Solution_ABCD_EFGH}
\end{table*}

\begin{figure*}
\centering
\includegraphics[width=0.36\textwidth]{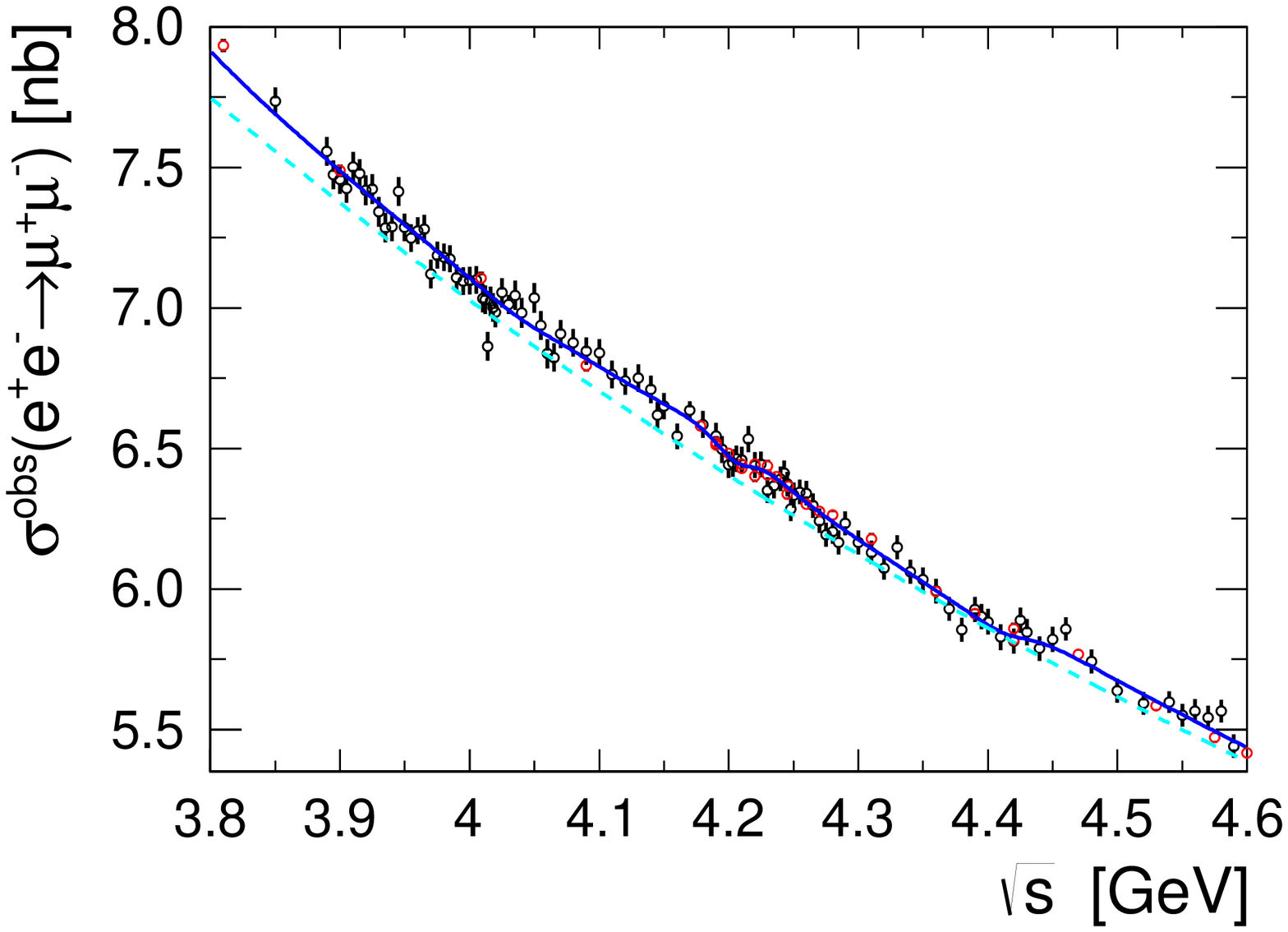}
\includegraphics[width=0.36\textwidth]{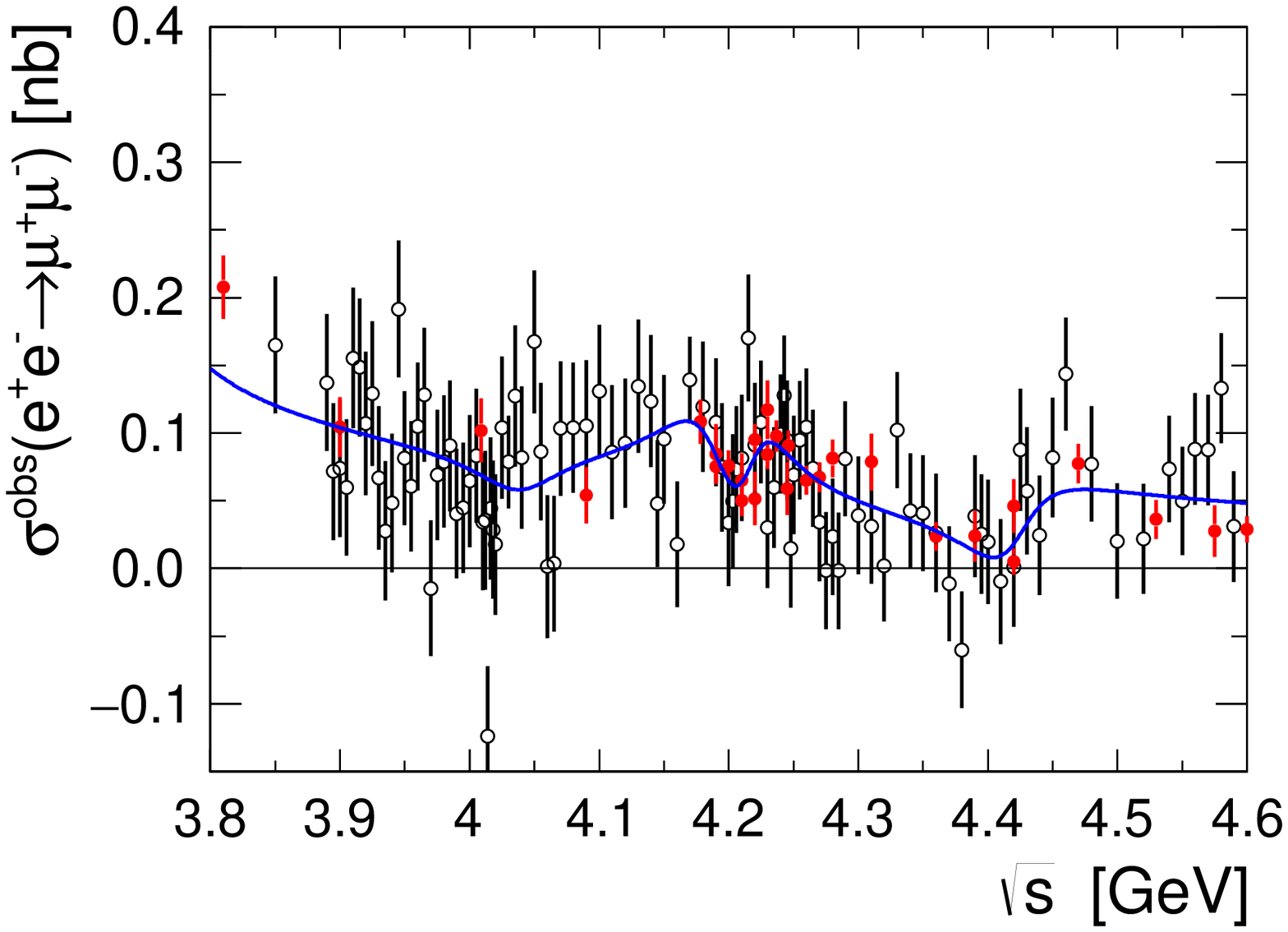}
\caption{
Measured cross sections for $e^+e^- \rightarrow \mu^+\mu^-$ with the
fit superimposed. The left plot shows the absolute cross sections,
while the right plot shows the cross section after subtraction of both
the continuum and  $\psi(3686)\rightarrow\mu^+\mu^-$ contributions
(see text for details).
  }
\label{fig:cp4040p4160s42zzp4415_124p_oknewindcntp_24july2018_v1a_RscanData}
\end{figure*}

Figure~\ref{fig:cp4040p4160s42zzp4415_124p_oknewindcntp_24july2018_v1a_RscanData}
(left)
shows the observed cross sections
with a fit to the sum of eleven contributions:
continuum $e^+e^-\rightarrow \mu^+\mu^-$,
the nine known vector states and the $S(4220)$ decay into $\mu^+\mu^-$.
The black empty circles
in Fig.~\ref{fig:cp4040p4160s42zzp4415_124p_oknewindcntp_24july2018_v1a_RscanData}
are for the lower luminosity data
(integrated luminosity less than 12 pb$^{-1}$),
the filled red circles are for the
higher luminosity data,
the solid line is for the fit, and the dashed line is for the contribution
from the  $e^+e^-\rightarrow \mu^+\mu^-$ continuum.
Figure~\ref{fig:cp4040p4160s42zzp4415_124p_oknewindcntp_24july2018_v1a_RscanData}
(right)
shows the corresponding observed cross section, for which both the contributions from the continuum
$e^+e^-\rightarrow \mu^+\mu^-$ and the decay $\psi(3686)\rightarrow \mu^+\mu^-$ are
subtracted.
Removing the $S(4220)$ from the fit yields a $\chi^2$ change by 23.78, for a reduction of
four degrees of freedom, which corresponds to a statistical
significance for the structure of $3.9\sigma$.

The systematic uncertainties on the values of the parameters given in
Table~\ref{tab:FitResults_CntPsi3686S37xxP3770P4040P4160S4230P4415_Solution_ABCD_EFGH}
originate from three sources: (1) systematic uncertainties on the
observed cross sections, (2) uncertainties on the parameters for the
$\psi(3686)$, $\psi(3770)$, $\psi(4040)$, $\psi(4160)$, and $\psi(4415)$ states,
(3) uncertainties on the c.m.\ energies.
Adding these contributions in quadrature we obtain the total systematic uncertainties
for the fit parameters, which are listed as the second uncertainties
in
Table~\ref{tab:FitResults_CntPsi3686S37xxP3770P4040P4160S4230P4415_Solution_ABCD_EFGH}.

\begin{figure*}
\centering
\includegraphics[width=0.36\textwidth]{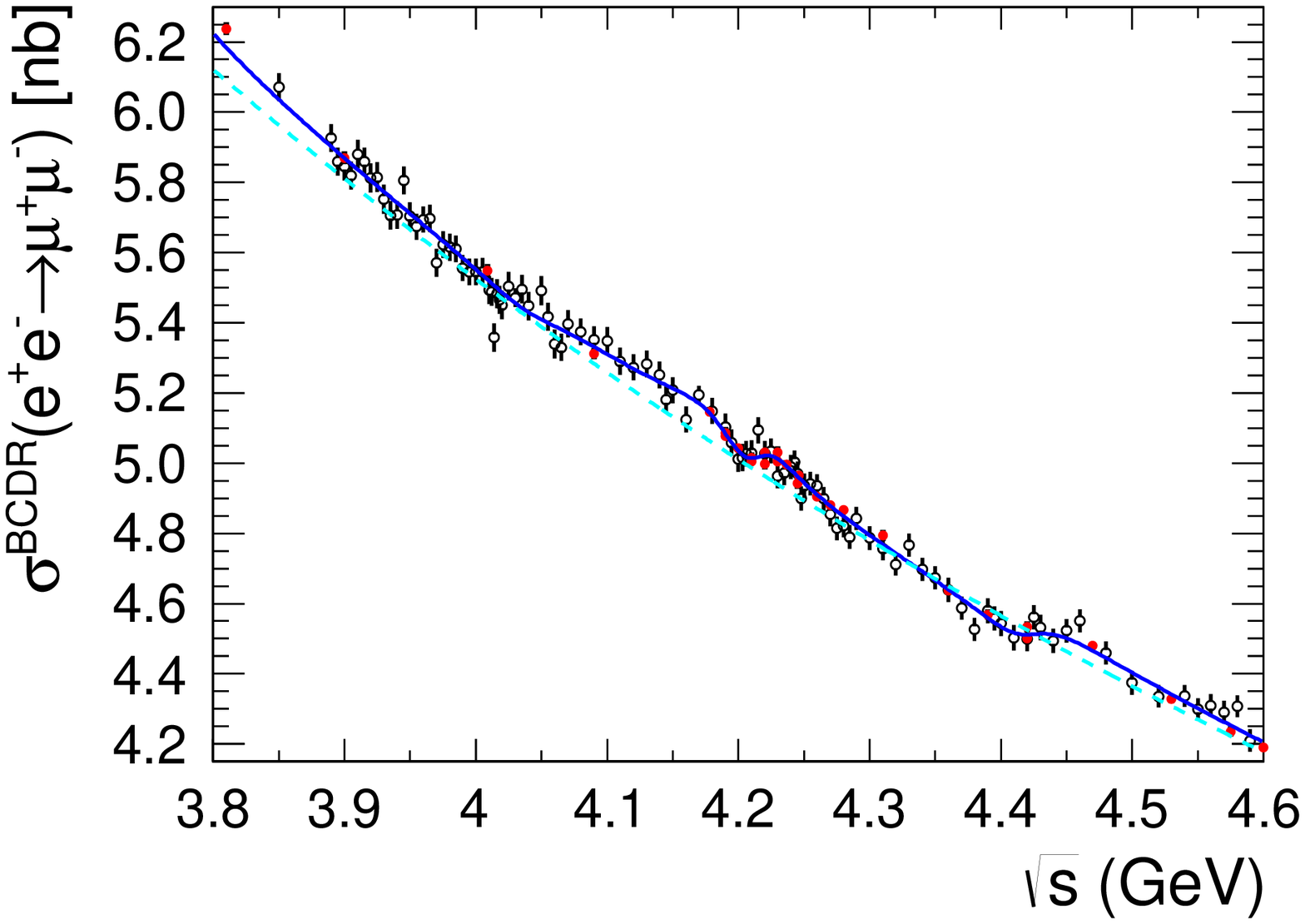}
\includegraphics[width=0.36\textwidth]{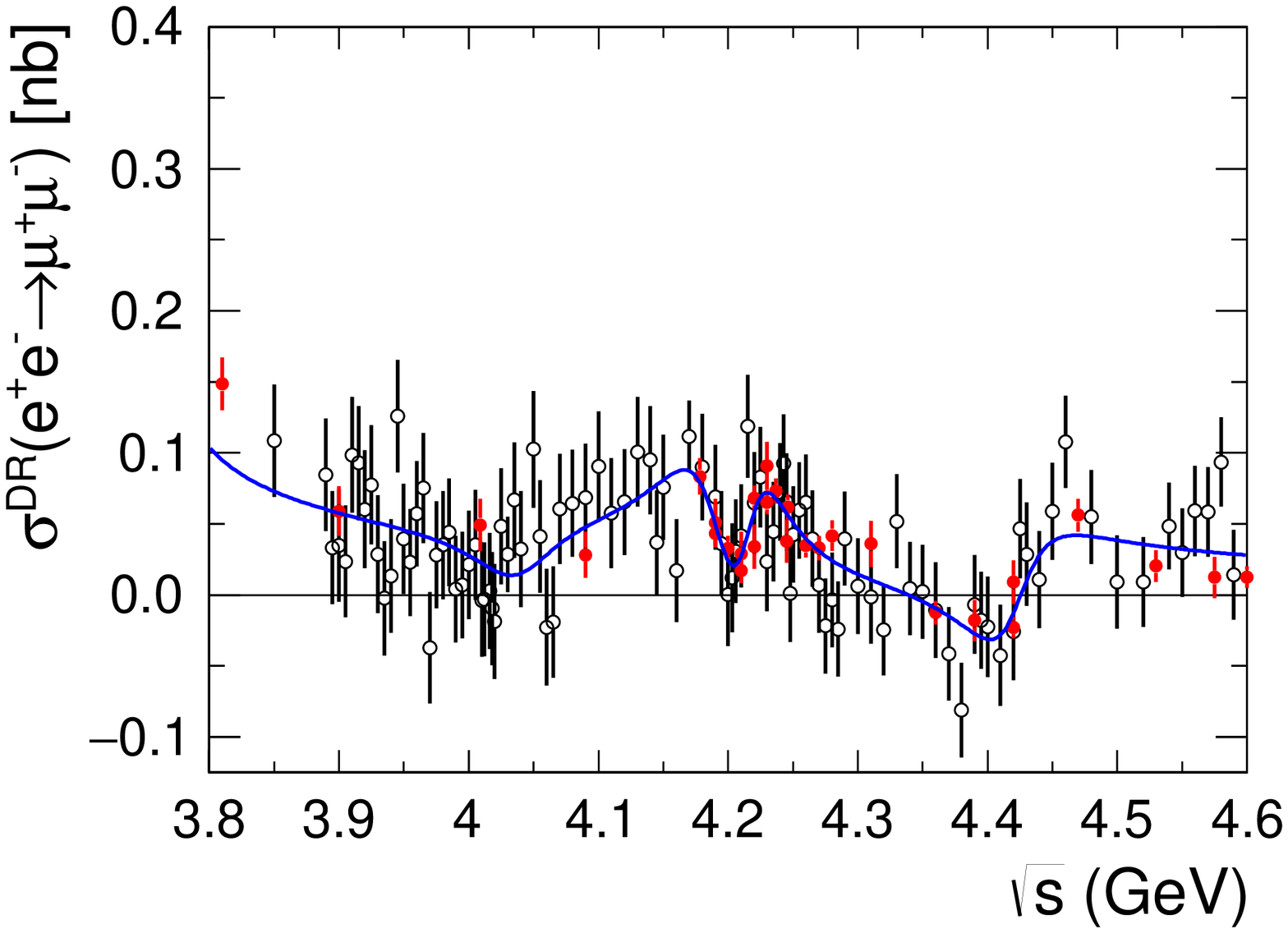}
\caption{
Corresponding BCDR cross sections for $e^+e^- \rightarrow \mu^+\mu^-$ with the
fit superimposed. The left plot shows the absolute cross sections, while the right has
subtracted both the continuum and  $\psi(3686)\rightarrow\mu^+\mu^-$ contributions (see text for details).
  }
\label{fig:cp4040p4160s42zzp4415_124p_oknewindcntp_24july2018_v1a_RscanData_BCDR}
\end{figure*}

Initial State Radiation distorts the shape of the
resonances in the observed cross section.
Most ISR events not only populate the valleys between the resonance peaks
(see cross section around 4.02, 4.20, and 4.36 GeV in
Fig.~\ref{fig:cp4040p4160s42zzp4415_124p_oknewindcntp_24july2018_v1a_RscanData} (right)),
but also reduce the heights of these peaks,
which weakens the significance of the signals.
Figure~\ref{fig:cp4040p4160s42zzp4415_124p_oknewindcntp_24july2018_v1a_RscanData_BCDR}
(left)
shows the corresponding Born-continuum-dressed-resonance (BCDR) cross section, which is
the observed cross section divided by the ISR correction factor
$f_{\rm ISR}(s)$, with
$f_{\rm ISR}(s)=\sigma^{\rm obs}_{\mu^+\mu^-}(s)/\sigma^{\rm D}_{\mu^+\mu^-}(s)$,
where $\sigma^{\rm obs}_{\mu^+\mu^-}(s)$ is given
in Eq.~(\ref{equation:Expected_CrossSection})
and  $\sigma^{\rm D}_{\mu^+\mu^-}(s)$
is given in Eq.~(\ref{equation:Drss_CrossSection_A}) with $x=0$.
The BCDR cross section is the sum of the Born continuum cross
section of $e^+e^-\rightarrow \mu^+\mu^-$ and the dressed cross sections for
the resonances decaying into $\mu^+\mu^-$.
The ISR correction
removes the ISR-return events
(see cross section around 4.02, 4.20, and 4.36 GeV in
Fig.\ref{fig:cp4040p4160s42zzp4415_124p_oknewindcntp_24july2018_v1a_RscanData_BCDR} (right))
and restores the heights of the signal peaks,
making the $S(4220)$ signal to be more pronounced 
and more clearly seen in the BCDR cross sections.
Removing the $S(4220)$ from the fit to the BCDR cross section causes 
a $\chi^2$ change by 78.20, for a reduction of four degrees of freedom.
This change corresponds to a statistical
significance of more than $7\sigma$ for the $S(4220)$ structure.
Analysis of an ensemble of simulated data sets of $e^+e^-\rightarrow h_c \pi^+\pi^-$ 
generated using the Y(4220) and Y(4390) resonance parameters~\cite{PRL118_092001_2017} 
demonstrates that the signal significance of structures seen in the dressed cross section 
typically exceeds those seen in the observed cross section by about 4 sigma, 
which is compatible with the increase seen in the data.

    The eight solutions summarized in
Table~\ref{tab:FitResults_CntPsi3686S37xxP3770P4040P4160S4230P4415_Solution_ABCD_EFGH}
have $\chi^2$ values 135.47, 135.71, 135.76,  135.48, 135.59, 135.95, 135.67, 135.61, 
respectively, for 121 degrees of freedom.
Thus, all of them are acceptable.
We choose  Solution I as the nominal result of the analysis.
The mass and total width of the $S(4220)$ determined from the fit
are consistent with those of the $Y(4220)$, $R(4230)$ and $R(4220)$ resonances
measured by the BESIII
Collaboration~\cite{Y4220_BESIII_PRL118_092002_2017,PRL114_092993_2015,PRL118_092001_2017},
so these are likely to be the same vector state.
With this assumption
we obtain the ratios of branching fractions:
 ${{\mathcal B(S(4220) \rightarrow \omega \chi_{c0} )}}/
   {\mathcal B(S(4220) \rightarrow \mu^+\mu^-)}  = 54\pm 77$,
 ${{\mathcal B(S(4220) \rightarrow h_c \pi^+\pi^- )}}/
             {\mathcal B(S(4220) \rightarrow \mu^+\mu^-)}=92^{+142}_{-133}$
and
 ${{\mathcal B(S(4220) \rightarrow J/\psi\pi^+\pi^-)}}/
    {\mathcal B(S(4220) \rightarrow \mu^+\mu^-)} =(32\pm 46)~{\rm to}~(266\pm 373)$,
where the uncertainties include both statistical and systematic contributions.
These ratios indicate
that the branching fraction
of the decay $S(4220) \rightarrow \mu^+\mu^-$  is typically two orders of magnitude
smaller than $S(4220)\rightarrow{M_{c\bar c}{\rm X_{LH}}}$ decays.

Our measured muonic widths for
the $\psi(4040)$
and $\psi(4415)$
are consistent within $\sim 1.3$ times the uncertainties
with theoretical expectations for the electronic widths of these states,
which are $1.42$ and  $0.70$~keV, respectively~\cite{PhysRevD79_094004_Y2009}.

In summary, we have measured the cross section
for $e^+e^- \rightarrow \mu^+\mu^-$ at c.m.\ energies from 3.8 to 4.6 GeV.
For the first time we have directly measured the muonic widths and branching fractions of
$\psi(4040)$, $\psi(4160)$ and $\psi(4415)$,
and determined the
phases of the decay amplitudes for these three resonances.
The relative phases of these three vector states range from
($-78\pm 33$) to ($157\pm 39$) degrees.

In addition, we have found evidence for a structure $S(4220)$ lying
near to 4.22 GeV/$c^2$ with a mass of
${M}_{S(4220)}=4216.7 \pm 8.9 \pm 4.1$~MeV/$c^2$,
a total width of ${\rm \Gamma^{\rm tot}_{S(4220)}}=47.2 \pm 22.8 \pm 10.5$~MeV,
and a muonic width of ${\rm \Gamma}^{\mu\mu}_{S(4220)}=1.53\pm1.26\pm0.54$~keV,
where the first uncertainties are statistical and the second are systematic.
The statistical significance of the $S(4220)$ signal is $3.9\sigma$.
The analysis of the BCDR cross section
of $e^+e^- \rightarrow \mu^+\mu^-$ yields
a statistical significance of the $S(4220)$ signal of more than 7$\sigma$.
Although the dimuon branching fractions of the $X_{\rm {above~D{\bar D}}}$ decays
are all at the level of $\sim 10^{-5}$,
the interference of these decays
with the $e^+e^- \rightarrow \mu^+\mu^-$ continuum
produces a measurable contribution to the cross section,
whose shape is sensitive to new states. Therefore the analysis of the $e^+e^-\to\mu^+\mu^-$ cross section
in the energy region between 3.73 and 4.8 GeV
is a promising way to discover new vector states, complementing the study of the processes
$e^+ e^- \rightarrow J/\psi X$ and $e^+ e^- \rightarrow {M_{c\bar c}X_{\rm LH}}$.

The BESIII collaboration thanks the staff of BEPCII and the IHEP computing
center for their strong support. This work is supported in part by National
Key Basic Research Program of China under Contract No.  
2015CB856700,
2009CB825204;
National Natural Science Foundation of China (NSFC) under Contracts Nos.
11625523, 11635010, 11735014, 11822506, 11835012, 11961141012, 
10935007;
the Chinese
Academy of Sciences (CAS) Large-Scale Scientific Facility Program; Joint
Large-Scale Scientific Facility Funds of the NSFC and CAS under Contracts
Nos. U1532257, U1532258, U1732263, U1832207; CAS Key Research Program of
Frontier Sciences under Contracts Nos. QYZDJ-SSW-SLH003, QYZDJ-SSW-SLH040;
100 Talents Program of CAS,
CAS Other Research Program under Code No. Y129360;
INPAC and Shanghai Key Laboratory for Particle
Physics and Cosmology; ERC under Contract No. 758462; German Research
Foundation DFG under Contracts Nos. Collaborative Research Center CRC 1044,
FOR 2359; Istituto Nazionale di Fisica Nucleare, Italy; Ministry of
Development of Turkey under Contract No. DPT2006K-120470; National Science
and Technology fund; STFC (United Kingdom); The Knut and Alice Wallenberg
Foundation (Sweden) under Contract No. 2016.0157; The Royal Society, UK
under Contracts Nos. DH140054, DH160214; The Swedish Research Council; U. S.
Department of Energy under Contracts Nos. DE-FG02-05ER41374, DE-SC-0010118,
DE-SC-0012069.

\end{document}